\documentclass[12pt]{article}
\usepackage{amssymb}
\usepackage{setspace}

\begin{document}

\newcommand{\bra}[1]{\langle #1|}
\newcommand{\ket}[1]{|#1\rangle}
\newcommand{\braket}[2]{\langle #1|#2\rangle}
\newcommand{\en}{\textrm{End}}
\newcommand{\g}{\theta}
\newcommand{\co}[2]{\lbrack #1 , #2 \rbrack}
\newcommand{\be}{\begin{eqnarray}}
\newcommand{\ee}{\end{eqnarray}}
\newcommand{\ad}{\textrm{ad}}
\newcommand{\p}{\partial}
\newcommand{\minint}{\textrm{$\int$}}

\def\maketitle{
   \begin{flushright}
   J. Math. Phys. \textbf{46}, 042308 (2005)\\
   DAMTP-2004-156
   \end{flushright}
   \begin{center}
   \Huge{On Extensions of Superconformal Algebras}\\
   \vskip 0.7 cm
   \large{Jasbir Nagi}\\ \vskip 0.7 cm \large{J.S.Nagi@damtp.cam.ac.uk}\\ \vskip 0.7 cm \large{DAMTP,
   University of Cambridge, Wilberforce Road,}\\ \vskip 0.7 cm \large{Cambridge,
    UK, CB3 0WA}
   \vskip 0.7 cm
   \end{center}
   }

\maketitle
\begin{abstract}
Starting from vector fields that preserve a differential form on a
Riemann sphere with Grassmann variables, one can
construct a Superconformal Algebra by considering central extensions
of the algebra of vector fields. In this note, the N=4 case is analyzed closely,
where the presence of weight zero operators in the field theory forces
the introduction of non-central extensions. How this modifies the
existing Field Theory, Representation Theory and Gelfand-Fuchs
constructions is discussed. It is also discussed how graded Riemann
sphere geometry can be used to give a geometrical description of the
central charge in the N=1 theory.
\end{abstract}
\newpage
\tableofcontents

\section{Introduction}

Two dimensional conformal symmetry in Quantum Field Theory has, over the last 30 years,
touched many parts of mathematics and theoretical physics. A Quantum
Field Theory that is conformally invariant is called a
Conformal Field Theory (CFT), and in this note, only the case of two
dimensions is examined. Lying at the heart of CFT is understanding how
to treat the infinite conformal symmetry on the quantum level, and
understanding the representation theory of the algebra. If one wants a
non-trivial, unitary representation of the symmetry algebra (known as the
DeWitt algebra), then a central extension must be introduced into the
algebra, yielding the Virasoro algebra\cite{kent}. Hence, when considering any
algebra which has the Virasoro algebra as a subalgebra, the
understanding of how and what extensions can be added is of crucial
importance. Two-dimensional CFTs present an example of a Quantum Field
Theory where
there is a rich interplay between the geometry of the theory and the
quantum theory. As a result, many aspects of the quantum theory can be
described elegantly by the geometry. This is a point of view that this
note will use repeatedly.

In this note, the case of the conformal symmetries of a Riemann sphere and
graded Riemann sphere are examined. In \cite{nagi} it was found that a
graded Riemann sphere is a sensible space on which to try and
construct a Superconformal Field Theory (SCFT). On this space many results can be
obtained, and primary fields (fields that `generate' the space of states) can be built in a natural manner
associated to the geometry of the space. These spaces give rise to lie graded
algebras of vector fields that contain the DeWitt algebra. Much work
has been done on the extensions of these algebras in
\cite{kac3}. Here, the $N=4$ case is revisited, with particular
attention paid to the weight $0$ fields that arise in the
theory. These fields give unusual behaviour, giving logarithms in the
super OPE, and exhibiting some Jordan block structure in the adjoint representation. The structure,
however, turns out to be quite manageable, since the behaviour turns
out to be quite similar to that of a free boson. The
understanding of how to construct bosonic primary fields geometrically, as
sections of a line bundle is extremely well covered in the CFT
literature. However, it is also known, in the bosonic theory, how to
construct central extensions geometrically\cite{witten}. Here, this is
extended to the $N=1$ case, and is discussed how this might extend to
higher $N$.

In Section 2, the vector fields for the $N=4$ case are found, and
extensions of the algebra considered, using just the graded Jacobi
Identity. In particular, it is found that if the algebra is not
reduced to its simple subalgebra, then the Jacobi identity implies
that algebra must be extended by
non-central elements, if it is to contain the Virasoro algebra with
non-trivial central charge. Section 3 examines how this fits into the
operator formalism of CFT, where the starting point is usually an
Operator Product expansion. There the subtlety arises from
understanding what mode expansions to take for operators, and how to
treat logarithms in the Operator Product expansion. Section 4 looks at how the usual
representation theory of the $N=4$ algebra will be altered with
the non-central extension found. Section 5 then considers how the
algebra obtained fits into the formalism of Gelfand-Fuchs extensions,
and how superspace techniques can be used to write them. Section 6
then looks at the $N=1$ Gelfand-Fuchs cocycle, i.e. the $N=1$ central charge, and considers how to
realise the cocycle as a geometric object on a graded Riemann sphere.

\section{The algebra of Vector Fields}
On a Riemann sphere, one obtains the the DeWitt algebra by looking at
the vector fields that preserve the one-form $dz$, a basis of which is
given by
\be
l_n = -z^{n+1}\frac{\p}{\p z}
\ee
Calculating the commutation relations given by the lie bracket yields
the DeWitt algebra
\be
\co{l_n}{l_m} = (n-m)l_{n+m}
\ee
Similarly, on a graded Riemann sphere, with $N=4$ and the usual one-form $\omega
= dz + \sum_{i=1}^4\g_i d\g_i$, one finds a basis of vector fields that preserve $\omega$\cite{kac3}
\be\label{n4vec}
l_n &=& -z^n\Big( z\partial + \frac{1}{2}(n+1)\theta_i\partial_i\Big)
\nonumber\\
g^j_r &=& z^{r+\frac{1}{2}}\Big( \g_j\partial - \partial_j +
\frac{(r+\frac{1}{2})}{z}\g_j\g_k\partial_k\Big)\nonumber\\
t^{ml}_n &=& z^n\Big( \g_l\partial_m - \g_m\partial_l +
\frac{n}{z}\g_m\g_l\g_p\partial_p\Big)\nonumber\\
\psi_r^k &=&
-z^{r-\frac{1}{2}}\Big(\frac{1}{6}\epsilon_{kpqr}\g_p\g_q\g_r\partial
+ \frac{(r-\frac{1}{2})}{z}\g_1\g_2\g_3\g_4\partial_k +
\frac{1}{2}\epsilon_{kpqr}\g_p\g_q\partial_r\Big)\nonumber\\
u_n &=& -z^{n-1}\Big(\g_1\g_2\g_3\g_4\partial +
\frac{1}{12}\epsilon_{ijkl}\g_i\g_j\g_k\partial_l\Big)
\ee
These vector fields give rise to the graded commutation relations
under the graded Lie bracket
\be
&\co{l_n}{l_m}=(n-m)l_{m+n}\qquad\co{l_n}{g^j_{r}}=(\frac{n}{2}-r)g^j_{n+r}\qquad\co{l_n}{t^{pq}_m}=-mt^{pq}_m\nonumber\\&\nonumber\\&
\co{l_n}{\psi^k_r}=(-\frac{n}{2}-r)\psi^k_{n+r}\qquad\co{l_n}{u_m}=-(n+m)u_{m+n}\nonumber\\&\nonumber\\&
\co{g^j_r}{g^k_s}=2\delta_{jk}l_{r+s}+(s-r)t^{jk}_{r+s}\qquad
\co{t^{ml}_n}{g^j_r} = \delta_{mj}g^l_{n+r}-\delta_{lj}g^m_{n+r}+
n\epsilon_{mljk}\psi^k_{n+r}\nonumber\\\nonumber\\&
\co{g^j_r}{\psi^k_s}=\delta_{jk}2(r+s)u_{r+s} +
\frac{1}{2}\epsilon_{jkpq}t^{pq}_{r+s}\qquad\co{g^j_r}{u_n}=-\frac{1}{2}\psi^j_{n+r}\nonumber\\&\nonumber\\&
\co{t^{ml}_n}{t^{pq}_r} = \delta_{mq}t^{pl}_{n+r} +
\delta_{mp}t^{lq}_{n+r} + \delta_{lq}t^{mp}_{n+r} +
\delta_{lp}t^{qm}_{n+r}\quad\co{t^{ml}_n}{u_p}=0\nonumber\\&\nonumber\\&
\co{t^{ml}_n}{\psi^k_r}=\delta_{mk}\psi^l_{n+r}-\delta_{kl}\psi^m_{n+r}\qquad
\co{\psi^k_r}{\psi^j_s}=0\qquad\co{\psi^k_r}{u_n}=0\qquad \co{u_m}{u_n}=0
\ee
Note that this algebra is not simple, and that
$\co{\mathfrak{g}}{\mathfrak{g}}=\mathfrak{g} / \langle u_o\rangle$
(quotient taken in the vector space sense) is
simple. This simple algebra is the large $N=4$ algebra, without its
three central elements. The $t_n^{ml}$ form an $so(4)$ loop algebra. This loop algebra may be written
explicitly as $so(4)=su(2)\oplus su(2)$ by the change of basis
\be
&t^1_m = \frac{1}{2}(t_m^{12}+t_m^{34}) \qquad t^2_m =
\frac{1}{2}(t_m^{13}-t_m^{24}) \qquad t^3_m =
\frac{1}{2}(t_m^{14}+t_m^{23})\nonumber\\ &\nonumber\\
&\bar{t}^1_m = \frac{1}{2}(t_m^{34}-t_m^{12}) \qquad \bar{t}^2_m =
\frac{1}{2}(-t_m^{13}-t_m^{24}) \qquad \bar{t}^3_m = \frac{1}{2}(t_m^{23}-t_m^{14})\nonumber
\ee
These $su(2)$s can then be centrally extended, to affine currents,
with extension $c$ and $\bar{c}$. Defining $c^+ = c+\bar{c}$ and $c^-
= c-\bar{c}$, one finds the modified commutation relation
\be
\co{t^{ml}_n}{t^{pq}_r} &=& \delta_{mq}t^{pl}_{n+r} +
\delta_{mp}t^{lq}_{n+r} + \delta_{lq}t^{mp}_{n+r} +
\delta_{lp}t^{qm}_{n+r} \nonumber\\&&+(\delta_{mp}\delta_{lq}-\delta_{mq}\delta_{lp})c^+n\delta_{n+r,0} + \epsilon_{mlpq}c^-n\delta_{n+r,0}\nonumber
\ee
The $c^-$ also modifies, by the Jacobi identity, the relations
\be
&\co{g^j_r}{\psi^k_s}=\delta_{jk}2(r+s)u_{r+s} +
\frac{1}{2}\epsilon_{jkpq}t^{pq}_{r+s} +
c^-(r+\frac{1}{2})\delta_{jk}\delta_{r+s,0}\nonumber\\&\nonumber\\&
\co{l_n}{u_m}=-(n+m)u_{m+n} - \frac{c^-}{4}(n+1)\delta_{m+n,0}\nonumber
\ee
The $c^+$ also modifies, by the Jacobi identity, the relations
\be
&\co{l_m}{l_n} = (m-n)l_{m+n} - \frac{c^+}{4}m(m^2-1)\delta_{m+n,0}\nonumber\\&\nonumber\\&
\co{g^j_r}{g^k_s} = 2\delta_{jk}l_{r+s} + (s-r)t^{jk}_{r+s} -
c^+(r^2-\frac{1}{4})\delta_{jk}\delta_{r+s,0}\nonumber\\&\nonumber\\&
\co{\psi^k_r}{\psi^j_s} = c^+\delta_{jk}\delta_{r+s,0}\qquad
\co{u_m}{u_n} = -\frac{c^+}{4m}\delta_{m+n,0}\textrm{ for }m\neq 0\nonumber
\ee
As it stands, the $\{ g^j_r, \psi^j_s, u_0\}$ Jacobi identity implies
$c^+=0$. This offending Jacobi identity is usually bypassed by working
in $\co{\mathfrak{g}}{\mathfrak{g}}$ rather than in $\mathfrak{g}$, but
this is not the route that will be taken here. Non-zero $c^+$
can be obtained by adding another extension, denoted $v_0$. From the
$\co{l_n}{u_m}$ commutator, it can be seen that the $u_m$ form a current of weight zero. This current can be deformed
to include a logarithmic term, so that $u(z) = -\sum_n u_n z^{-n} + v_0
\log z$. This then modifies the commutation relations
\be
&\co{l_n}{u_m} = -(m+n)u_{m+n} -
v_0\delta_{m+n,0}\nonumber\\&\nonumber\\&\co{g_r^j}{\psi_s^k}= 2\delta_{jk}\big(
(r+s)u_{r+s} + v_0\delta_{r+s,0} \big) + \frac{1}{2}\epsilon_{jkpq}t^{pq}_{r+s}
\ee
Using the Jacobi identity, one can see that $v_0$ commutes with all
elements, except $u_0$. The $\{ g^j_r, \psi^j_s, u_0\}$ Jacobi
identity now yields $\co{u_0}{v_0}=\frac{c^+}{4}$. This
algebra now realises centrally extended $K'(4)$ \cite{kac3} (also known as large $N=4$\cite{vanp}) with the $u_0$
operator put back in. Note that in $K'(4)$, $v_0$ is a central
extension\cite{kac3}. The behaviour of the $u_m, v_0$ is very similar to that of
the modes of a free boson, identifying $v_0$ with momentum and $u_0$
with position. The commutation relations then become
\be\label{commrelns}
&\co{l_m}{l_n}=(m-n)l_{m+n}-\frac{c^+}{4}m(m^2-1)\delta_{m+n,0}\qquad\co{l_n}{g^j_{r}}=(\frac{n}{2}-r)g^j_{n+r}\nonumber\\&\nonumber\\&
\co{l_n}{\psi^k_r}=(-\frac{n}{2}-r)\psi^k_{n+r}\qquad\co{l_n}{u_m}=-(n+m)u_{m+n}-\big( \frac{c^-}{4}(n+1)
+v_0\big) \delta_{m+n,0}\nonumber\\&\nonumber\\&
\co{g^j_r}{g^k_s}=2\delta_{jk}l_{r+s}+(s-r)t^{jk}_{r+s} - c^+(r^2-\frac{1}{4})\delta_{jk}\delta_{r+s,0}\qquad\co{l_n}{t^{pq}_m}=-mt^{pq}_m\nonumber\\&\nonumber\\&
\co{t^{ml}_n}{g^j_r} = \delta_{mj}g^l_{n+r}-\delta_{lj}g^m_{n+r}+
n\epsilon_{mljk}\psi^k_{n+r}\qquad\co{g^j_r}{u_n}=-\frac{1}{2}\psi^j_{n+r}\nonumber\\\nonumber\\&
\co{g^j_r}{\psi^k_s}=\delta_{jk}2(r+s)u_{r+s} +
\frac{1}{2}\epsilon_{jkpq}t^{pq}_{r+s} + \big( c^-(r+\frac{1}{2})+2v_0\big)
\delta_{jk}\delta_{r+s,0}\nonumber\\&\nonumber\\&
\co{t^{ml}_n}{t^{pq}_r} = \delta_{mq}t^{pl}_{n+r} +
\delta_{mp}t^{lq}_{n+r} + \delta_{lq}t^{mp}_{n+r} +
\delta_{lp}t^{qm}_{n+r}\nonumber\\&\qquad\qquad\qquad\qquad+(\delta_{mp}\delta_{lq}-\delta_{mq}\delta_{lp})c^+n\delta_{n+r,0}
+ \epsilon_{mlpq}c^-n\delta_{n+r,0}\nonumber\\&\nonumber\\&
\co{t^{ml}_n}{u_p}=0\qquad
\co{t^{ml}_n}{\psi^k_r}=\delta_{mk}\psi^l_{n+r}-\delta_{kl}\psi^m_{n+r}\qquad
\co{\psi^k_r}{\psi^j_s}=c^+\delta_{jk}\delta_{r+s,0}\nonumber\\&\nonumber\\&\co{\psi^k_r}{u_n}=0\qquad\co{u_m}{u_n}=-\frac{c^+}{4m}\delta_{m+n,0} \qquad
\co{u_0}{v_0}=\frac{c^+}{4}
\ee
with $c^\pm$ central, and $v_0$ has only one non-trivial commutator,
namely $\co{u_0}{v_0}$. Thus, whilst $u_0, v_0$ can both be considered to be
operators at level $0$, they
cannot both be in the Cartan subalgebra. If $v_0$ is non-zero, one can
choose it to be in the Cartan subalgebra. Usually, in a Conformal
Field Theory, one finds that the space of operators at level zero can be identified with the Cartan subalgebra. This is not
the case here, and can potentially lead to Jordan Blocks. In this sense, the $N=4$ theory can be
thought of as a logarithmic theory, with the logarithmic character coming
from $u(z)$. The usual large algebra comes from looking at the simple
subalgebra obtained from identifying $u_0 \sim 0$, i.e. considering
the field $\partial u(z)$ as fundamental rather than $u(z)$. To see where the logarithms actually
come in, one must look at the operator formalism.

\section{The Operator Approach}

The super Operator Product Expansion of the super Virasoro operator
with a primary superfield for the $N=4$ case is given by\cite{ivan}
\be
\mathbb{T}(Z_1)\Phi(Z_2) &\sim&
\frac{h\g_{12,1}\g_{12,2}\g_{12,3}\g_{12,4}}{Z_{12}^2}\Phi(Z_2) +
\frac{\g_{12,1}\g_{12,2}\g_{12,3}\g_{12,4}}{Z_{12}}\partial\Phi(Z_2)
\nonumber\\&&+
\frac{1}{12}\frac{\epsilon_{ijkl}\g_{12,i}\g_{12,j}\g_{12,k}}{Z_{12}}D_l\Phi(Z_2)
+
\frac{1}{4}\frac{\epsilon_{ijkl}\g_{12,i}\g_{12,j}J^{kl}}{Z_{12}}\Phi(Z_2)\nonumber\\&&\nonumber\\&&
+ p\log(Z_{12})\Phi(Z_2)
\ee
where $Z_1=(w,\chi_i)$, $Z_2=(z,\g_i)$, $\g_{12,i}=(\chi_i-\g_i)$,
$Z_{12}=(w-z-\chi_i\g_i)$,
\be\label{Texpan}
\mathbb{T}(Z_2)=\g_1\g_2\g_3\g_4 L(z) +
\frac{1}{12}\epsilon_{ijkl}\g_i\g_j\g_kG^l(z) +
\frac{1}{8}\epsilon_{ijkl}\g_i\g_jT^{kl}(z) + \frac{1}{2}\g_k\psi^k(z) - U(z)
\ee
and the $J^{ab}$ form an $so(4)$ algebra with commutation relations
given by $\frac{1}{2}t_0^{ab}$. $\log(Z_{12})$ is defined by
\be\label{logexpn}
\log(Z_{12}) = \log(w-z) - \sum_{p=1}^4 \frac{1}{p}\Big(\frac{\chi_i\g_i}{(w-z)}\Big)^p
\ee
The $\chi_i$ components of each side can be taken, giving
\be\label{OPEcommrelns}
L(w)\Phi(Z_2)&\sim&\Big(\frac{h}{(w-z)^2}+\frac{1}{(w-z)}\partial+\frac{1}{2(w-z)^2}\theta_i\partial_i
- \frac{\g_i\g_j J^{ij}}{(w-z)^2}\nonumber\\&&\qquad -
\frac{6\g_1\g_2\g_3\g_4}{(w-z)^4}p\Big)\Phi(Z_2)\nonumber\\&&\nonumber\\
\frac{1}{2}G^i(w)\Phi(Z_2)&\sim&\Big(\frac{-h\g_i}{(w-z)^2}-\frac{\g_i\partial}{2(w-z)}
+ \frac{\partial_i}{2(w-z)} -
\frac{\g_i\g_j\partial_j}{2(w-z)^2} \nonumber\\&& \qquad+ \frac{\g_i\g_j\g_k
  J^{jk}}{(w-z)^3} + \frac{\g_j J^{ji}}{(w-z)^2}-
\frac{\epsilon_{ijkl}\g_j\g_k\g_l}{3(w-z)^3}p\Big)\Phi(Z_2)\nonumber\\&&\nonumber\\
\frac{1}{2}T^{ab}(w)\Phi(Z_2)&\sim&\Big(\frac{h\g_a\g_b}{(w-z)^2} +
\frac{\g_b\partial_a-\g_a\partial_b}{2(w-z)} +
\frac{\g_a\g_b\g_j\partial_j}{2(w-z)^2} -
\frac{\g_1\g_2\g_3\g_4}{(w-z)^3}\epsilon_{abjk}J^{jk} \nonumber\\&&\qquad
+\frac{1}{(w-z)^2}(\g_a\g_jJ^{bj}-\g_b\g_jJ^{aj}) +
\frac{1}{(w-z)}J^{ab} \nonumber\\&&\qquad +
\frac{\epsilon_{abjk}\g_j\g_k}{2(w-z)^2}p\Big)\Phi(Z_2)\nonumber\\ &&\nonumber\\
\frac{1}{2}\psi^k(w)\Phi(Z_2)&\sim& \Big(
\frac{-h\epsilon_{klmn}\g_l\g_m\g_n}{6(w-z)^2} +
\frac{\epsilon_{klmn}\g_l\g_m\g_n}{12(w-z)}\partial +
\frac{\g_1\g_2\g_3\g_4}{2(w-z)^2}\partial_k +
\frac{\epsilon_{klmn}\g_l\g_m}{4(w-z)}\partial_n \nonumber\\&&\qquad-
\frac{\epsilon_{klmn}\g_l}{2(w-z)}J^{mn} +
\frac{\g_k\epsilon_{lmnp}\g_l\g_m}{4(w-z)^2}J^{np} -
\frac{\g_k}{(w-z)}p\Big)\Phi(Z_2)\nonumber\\&&\nonumber\\
-U(w)\Phi(Z_2)&\sim& \Big(\frac{h\g_1\g_2\g_3\g_4}{(w-z)^2} -
\frac{\g_1\g_2\g_3\g_4}{(w-z)}\partial -
\frac{\epsilon_{klmn}\g_k\g_l\g_m}{12(w-z)}\partial_n +
\frac{\epsilon_{klmn}\g_k\g_l}{4(w-z)}J^{mn}\nonumber\\&&\qquad + \log(w-z)p\Big)\Phi(Z_2) 
\ee
from which the vector fields of (\ref{n4vec})
can be recovered. Note that logarithms only appear in
OPEs containing $U(z)$. Clearly, in this last OPE, a contour integral can only be taken
if $\partial U(w)\Phi(z)$ is considered. Taking
$U(z)=\sum_nU_nz^{-n}+V_0\log z$, it can be shown that
$\co{V_0}{\Phi}=p\Phi$. Allowing $V_0$ to annihilate the
vacuum \cite{ft1} then
yields $V_0\ket{\Phi} = p\ket{\Phi}$. The $\co{V_0}{\Phi}$ commutator
is unusual, in that it contains no differential operators. Hence, it
is not obvious how to associate a conformal vector field of the form of
(\ref{n4vec}) to $V_0$. The
logarithm in the last OPE of (\ref{OPEcommrelns})
prevents one from obtaining an action from $U_0$. If one looks at a
representation where $V_0=0$, then it can be seen that $p=0$ and
that the $w$ contour integral in $U(w)\Phi(z)$ can be
performed, to give. 
\be
\co{U_0}{\Phi(Z)} \sim \Big( \frac{h\g_1\g_2\g_3\g_4}{z^2} +
\frac{\g_1\g_2\g_3\g_4\partial}{z} +
\frac{\epsilon_{klmn}\g_k\g_l\g_m\partial_n}{12z} - \frac{\epsilon_{klmn}\g_k\g_lJ^{mn}}{4z}\Big)\Phi(Z)\nonumber
\ee
However, its action on the highest weight from this approach is
unclear, and a more careful approach to the representation theory is warranted.

The logarithmic character can also be examined by looking at the
$\mathbb{T}(Z_1)\mathbb{T}(Z_2)$ OPE, given by
\be\label{ttope}
\mathbb{T}(Z_1)\mathbb{T}(Z_2)&\sim& \frac{c^+\log(Z_{12})}{4} -
\frac{c^-\g_{12,1}\g_{12,2}\g_{12,3}\g_{12,4}}{4Z_{12}^2} +
\frac{\g_{12,1}\g_{12,2}\g_{12,3}\g_{12,4}}{Z_{12}}\partial\mathbb{T}(Z_2)
+\nonumber\\ &&\qquad \frac{1}{12}\frac{\epsilon_{ijkl}\g_{12,i}\g_{12,j}\g_{12,k}}{Z_{12}}D_l\mathbb{T}(Z_2)
\ee
Using (\ref{logexpn}), one can see that the only term involving
logarithms is the term
\be\label{Ulog}
U(w)U(z) \sim \frac{c^+}{4}\log(w-z)
\ee
Therefore, if $c^+\neq 0$, $U(z)$ must have a logarithmic component in its mode
expansion. This can easily be verified by taking the contour integrals
in (\ref{Ulog}) to get the commutation relations. If there is no logarithmic
component in $U(z)$, then from the computation one can deduce that
$c^+ = 0$, the same result found when using the Jacobi identity in the
previous section. The
field $U(z)$ behaves in a very similar way to a free
boson. In particular, $\co{V_0}{U_0}=\frac{c^+}{4}$, and hence they
are not mutually diaganolizable, as was reflected by the above
manipulations of the last OPE in (\ref{OPEcommrelns}).

One can ask if the field, $U(z)$, can be written in a
logarithmic form \cite{flo}
\be
L(w)D(z) &\sim& \frac{hD(z) + E(z)}{(z-w)^2} + \frac{\partial
  D(z)}{(w-z)}\nonumber\\
L(w)E(z) &\sim& \frac{hE(z)}{(z-w)^2} + \frac{\partial
  E(z)}{(w-z)}\nonumber
\ee
From (\ref{ttope}) one can read off the $L(w)U(z)$ OPE to find
\be
L(w)U(z) \sim \frac{\partial U(z)}{(w-z)} - \frac{c^-}{4(w-z)^2}
\ee
Regarding $-\frac{c^-}{4}$ as a constant field of weight zero, one then
gets the desired form. Whilst the $V_0$ operator gives rise to an
eigenvalue, this analysis does not yield a conformal transformation
associated to $V_0$. The previous analysis shows that $v_0$ had to be
introduced as an extension of the algebra. This suggests that rather
than thinking of
the $p$ eigenvalue as being associated to a primary field, one should
instead think of the $V_0$ operator as
appearing in a similar way to a central extension.

\section{A little Representation Theory}

A closer look at the level zero operators is warranted. These are
normally defined as those operators with $\ad l_0$ eigenvalue being
zero. Considering the commutation relations (\ref{commrelns}), one can
see that clearly $\{l_0, t_0^{ml}, v_0, c^\pm \}$ fall into this
category. However, $u_0$ has a strange action under $\ad l_0$, namely
\be
\co{l_0}{u_0} = -\left(\frac{c^-}{4} + v_0\right) \nonumber
\ee
If one were to define a basis $e_1 = -\frac{c^-}{4} - v_0$, $e_2 = u_0$
and write down the matrix for $\ad l_0$ with respect to this basis,
one would find the Jordan block
\be
(\ad l_0) = \left( \begin{array}{cc} 0 & 1 \\ 0 & 0 \end{array}
\right) \nonumber
\ee
In this manner, $u_0$ can be considered as an operator at level
zero that is not in the Cartan subalgebra.

The algebra here differs slightly from the usual large $N=4$
algebra, by the mode $u_0$. This affects the representation
theory\cite{tao1}\cite{tao2}. The Fock space will be enlargened, due to the presence of
polynomials in $u_0$ acting on the highest weight. Using the analogy of $v_0$ as momentum, and $u_0$ as position,
instead of considering the states $u_0^n\ket{h}$, the `momentum' eigenstates
$\ket{k,h}=e^{-ku_0}\ket{h}$ can be considered. From the fact that the only
non-trivial commutators $u_0$ has are with $l_n, g_r^j, v_0$, one can
show
\be
&v_0\ket{k,h} = (\frac{c^+k}{4}+p)\ket{k,h}, \qquad g_r^j\ket{k,h} =
e^{-ku_0}g_r^j\ket{h} + \frac{k}{2}\psi_r^j\ket{k,h}\nonumber\\ \nonumber\\
&l_n\ket{k,h} = \left\{ \begin{array}{ll} e^{-ku_0}l_n\ket{h} + knu_n\ket{k,h} &
  n\neq 0\\ \Big(h + k(\frac{c^-}{4}+p) + k^2\frac{c^+}{8}  \Big)\ket{k,h}& n=0\end{array}\right.
\ee
from which it can be seen that $\ket{k,h}$ obeys highest weight
conditions, with potentially different $v_0$ and $l_0$ eigenvalues
from $\ket{h}$. In analogy with a free boson, the $u_n$. $n>0$
annihilate the vacuum and the highest weight state and $v_0$ annihilates
the vacuum. For non-zero $c^+$, $u_0$ annihilates neither.

\section{Gelfand-Fuchs 2-cocycles}

For an algebra of vector fields, where a function can be associated to
each vector field, it is often useful to construct central extensions
by considering Gelfand-Fuchs 2-cocycles\cite{groz}. Here, superfield
formalisms are used, which yield similar results to \cite{groz}. For instance, in the bosonic
case, one has\cite{witten}
\be
&l_m = -z^{m+1}\frac{\p}{\p z},\qquad l(z) = z^{m+1}, \textrm{ then}\nonumber\\
&c(l_m, l_n) = \frac{1}{24\pi i}\oint_0 dz \left(\frac{\p^3}{\p z^3}
z^{m+1}\right) z^{n+1} = \frac{c}{12}m(m^2-1)\delta_{m+n,0}
\ee
or for a general polynomial $l(z)$
\be
c(l_{(1)}, l_{(2)}) = \frac{1}{24\pi i}\oint_0 dz l_{(1)}'''l_{(2)}
\ee
where the contour is a closed loop around the origin, say the unit
circle, beginning and ending at $z=1$.
Since the extensions are known for the $N=1,2,3,4$ algebras, they
can be put in Gelfand-Fuchs 2-cocycle form. For the $N=1$ case
\be\label{n1param}
& l_n = -z^n(z\p_z + \frac{1}{2}(n+1)\g \p_\g),\qquad l(z)=z^{n+1}\nonumber\\
& g_r = z^{r+\frac{1}{2}}(\p_\g - \g\p_z),\qquad g(z) = z^{r+\frac{1}{2}}
\ee
Defining the graded field $X_{(i)} = \frac{1}{2}l_{(i)} + \g
g_{(i)}$ with $l$ and $g$ graded even polynomials in $z, z^{-1}$, one finds the central extension is given by the 2-cocycle
\be\label{n1centr}
c(X_{(1)}, X_{(2)}) = \frac{1}{6\pi i}\oint_0 dzd\g (DX_{(1)}'')X_{(2)}
\ee
where $D = \p_\g + \g\p_z$, and, as usual, $\int d\g$ really means
$\frac{\p}{\p\g}$. Similarly, for $N=2$, the vector fields and associated fields are
\be
& l_n = -z^n(z\p_z + \frac{1}{2}(n+1)\g_i\p_{\g_i}), \qquad l(z) =
z^{m+1}\nonumber\\
& g_r^i = z^{r-\frac{1}{2}}(z\g_i\p_z - z\p_{\g_i} +
(r+\frac{1}{2})\g_i\g_j\p_{\g_j}), \qquad g^i(z) =
z^{r+\frac{1}{2}}\nonumber\\
& t_m = -z^m(\g_2\p_{\g_1} - \g_1\p_{\g_2}),\qquad t(z)=z^m
\ee
Introducing the graded field
\be
X_{(i)} = \frac{1}{2}l_{(i)} + \g_j g^j_{(i)} +
\g_1\g_2t_{(i)}
\ee
the 2-cocycle is given by
\be
c(X_{(1)}, X_{(2)})=\frac{1}{6\pi i}\oint_0dzd\g_2d\g_1(D_1D_2X_{(1)}')X_{(2)}
\ee
where $D_i = \p_{\g_i} + \g_i\p_z$. For $N=3$,
\be
& l_n = -z^n(z\p_z + \frac{1}{2}(n+1)\g_i\p_{\g_i}), \qquad
l(z)=z^{m+1}\nonumber\\
& g_r^i = z^{r-\frac{1}{2}}(z\g_i\p_z - z\p_{\g_i} +
(r+\frac{1}{2})\g_i\g_j\p_{\g_j}), \qquad g^i(z) =
z^{r+\frac{1}{2}}\nonumber\\
& t^i_m = z^{m-1}(z\epsilon_{ijk}\g_j\p_{\g_k} -
m\g_1\g_2\g_3\p_{\g_i}), \qquad t^i(z) = z^m\nonumber\\
&\psi_r = -z^{r-\frac{1}{2}}(\g_1\g_2\g_3\p_z +
\frac{1}{2}\epsilon_{ijk}\g_i\g_k\p_{\g_k}), \qquad \psi(z) =
z^{r-\frac{1}{2}}\\ \nonumber\\
&X_{(i)} = \frac{1}{2}l_{(i)} + \g_j g^j_{(i)} +
\frac{1}{2}\epsilon_{klm}\g_k\g_lt^m_{(i)} + \g_1\g_2\g_3\psi_{(i)}
\ee
\be
c(X_{(1)}, X_{(2)}) = \frac{1}{6\pi i}\oint_0dzd\g_3d\g_2d\g_1(D_1D_2D_3X_{(1)})X_{(2)}
\ee
Now, for $N=4$, $X_{(i)}$ is given by
\be
X_{(i)} = \frac{1}{2}l_{(i)} + \g_jg^j_{(i)} +
\frac{1}{2}\g_a\g_bt^{ab}_{(i)} -
\frac{1}{6}\epsilon_{abcd}\g_a\g_b\g_c\psi^d_{(i)} - \g_1\g_2\g_3\g_4\frac{1}{2}u_{(i)}
\ee
where $u_{(i)} = z^{n-1}$ corresponds to the vector $u_m$ in (\ref{n4vec}), and similarly for the
other fields in $X_{(i)}$. In the cases so far, given an $X_{(i)}$, a conformal vector field can
be obtained. It is worth considering what $X_{(i)}$ means in the
operator approach. To this end, recall the super stress-energy tensor
$\mathbb{T}$ from (\ref{Texpan}). From the OPE (\ref{ttope}), it can be
shown that $L(z)$ scales like a field of weight $2$, and hence expansion $L(z)
= \sum_m L_m z^{-m-2}$. Similarly, the other operators have scaling
dimensions - $G^i$ is $\frac{3}{2}$, $T^{ij}$ is $1$, $\psi^i$ is
$\frac{1}{2}$ and $U$ is $0$. In fact, the $G^i, T^{ij}, \psi^i$ are
primary fields. Now, rather than obtain the vector field associated to
$X_{(i)}$, the operator associated to it can be obtained by computing
\be
\frac{1}{\pi i}\oint_0 dzd\g_4d\g_3d\g_2d\g_1 X_{(i)}\mathbb{T}
\ee
A similar formula \cite{ft2} holds for the smaller $N$.
Since $V_0$ is a part of $\mathbb{T}$, one can ask how to obtain the
operator $V_0$ from the above integral, and see if it sheds light on how the
$\co{u_0}{v_0}=\frac{c^+}{4}$ commutator might be obtained. To this end, consider
the logarithmic part of
\be
\frac{1}{\pi i}\int_{1+i\epsilon}^{1-i\epsilon} dz \frac{1}{2}u(z)U(z) &=& \frac{1}{\pi i}\int_{1+i\epsilon}^{1-i\epsilon}
dz \frac{1}{2}u(z)V_0 \log(z) \nonumber\\ &=& -\frac{1}{2\pi i}\int_{1+i\epsilon}^{1-i\epsilon} dz
\frac{1}{z}V_0\minint u + \frac{1}{2\pi i}V_0\left[ \log(z) \minint u\right]_{1+i\epsilon}^{1-i\epsilon}
\ee
Concentrating on the
first part of the expression, it seems very suggestive to associate
the constant part of $\int u$, which arises as an integration
constant, to the algebra element $v_0$ (assuming $\int u$ is single
valued around the origin, i.e. $u$ has no $\frac{1}{z}$ term). This
turns out to be precisely what is needed to obtain the $c^+$ 2-cocycle.
\be\label{cpgf}
c^+(X_{(1)}, X_{(2)}) = \frac{-1}{2\pi i}\oint_0
dzd\g_4d\g_3d\g_2d\g_1(D_1D_2D_3D_4\minint X_{(1)})X_{(2)}
\ee
assuming the integrand is single valued around the origin. On
expanding out the $X_{(i)}$ and applying all the superderivatives, the
only component of $X_{(1)}$ that is actually integrated in $D_1D_2D_3D_4\int X_{(1)}$ is the
$u_{(1)}$ component. This
essentially means that $c^+(u_0, u_0)$ cannot be explicitly obtained,
since the integrand would have $\log$s in it. This, however, is not a
problem, since $\co{u_0}{u_0}=0$ by antisymmetry of the
commutator. Also, $c^+(v_0, v_0)$ is not obtained, but by the same argument is
clearly zero. Most importantly, if the integration constant is taken
to be $1$, then one can obtain $c^+(v_0, u_0) = -\frac{1}{4}$. The
$c^-$ cocycle can also be found,
\be
c^-(X_{(1)}, X_{(2)}) = -\frac{1}{2\pi i}\oint_0
dzd\g_4d\g_3d\g_2d\g_1 X_{(1)}'X_{(2)}
\ee
as well as an expression for the $v_0$ extension.
\be
v_0(X_{(1)}, X_{(2)}) = \frac{2}{\pi i}\oint_0 dzd\g_4d\g_3d\g_2d\g_1\Big(\frac{1}{z}(1-\frac{1}{2}\g_i\p_{\g_i})X_{(1)}\Big)X_{(2)}
\ee
All of the extensions here for all $N$ are consistent with the
operator formalism. Apart from $c^+$ in $N=4$, which has a problem
with logs, all the extensions can be obtained from the super OPE by
calculating
\be
-\frac{1}{4\pi}\oint_0 dzd\g_N\ldots d\g_1 \oint_z dw d\chi_N \ldots d\chi_1 X_{(1)}(Z_1)X_{(2)}(Z_2)\mathbb{T}(Z_1)\mathbb{T}(Z_2)
\ee
and give rise to the same formulae. The formulae also suggest that the
extensions should be described by a map from two vector fields into
something proportional to the `volume form'
\be\label{conj}
C : \mathcal{D}^1\mathcal{A}_N \times
\mathcal{D}^1\mathcal{A}_N\rightarrow dz\otimes_i\frac{\p}{\p\g_i}
\ee

\section{$N=1$ Gelfand-Fuchs 2-cocycle}

In (\ref{n1param}), a parameterization of a vector field in terms of a
field was written. One can try and explore how these fields are
related to a graded Riemann sphere. First, redefine $X =
\frac{1}{2}f(z) + g(z)\g$, where now $f$ and $g$ need not have a
defined parity, i.e. they are each a sum of an even part and an odd
part. Now, introduce the map
\be
l:X\mapsto -\Big(f(z)\frac{\p}{\p z} + \frac{1}{2}f'(z)\g D\Big) -
g(z)\Big( 2\g\frac{\p}{\p z}-D\Big)
\ee
Under this identification, (\ref{n1centr}) holds. In components
(\ref{n1centr}) now reads
\be
c(X_{(1)}, X_{(2)}) = \frac{1}{6\pi i}\oint
dz\Big(\frac{1}{4}f_{(1)}'''f_{(2)} + (-1)^{g_{(2)}}g_{(1)}''g_{(2)}\Big)
\ee
where \cite{ft3}
$(-1)^{g_{(2)}}g_{(2)} = (-1)^{g_{(2)0}}g_{(2)0} + (-1)^{g_{(2)1}} g_{(2)1} = g_{(2)0} - g_{(2)1}$ with $g_{(2)0}$
and $g_{(2)1}$ being respectively the even and odd parts of $g_{(2)}$.
Recall that any invertible superconformal transformation can be parameterized by
$\Phi : (z,\g)\mapsto (w, \pi)$
\be\label{scfparam}
w &=& w_0 + \g w_1 (w_0')^\frac{1}{2}\nonumber\\
\chi &=& w_1 + \g (w_0' + w_1w_1')^\frac{1}{2} 
\ee
where $w_0=w_0(z)$ is even and with body, and $w_1=w_1(z)$ is
odd. Restricting to invertible transformations, and using the
superconformal condition $Dw = \chi D\chi$, one can show that
\be
\Phi^*\left( \begin{array}{c}\frac{\p}{\p w}\\D_\chi\end{array}\right) =
\left(\begin{array}{cc}(D_\g\chi)^{-2} & -\Big(\frac{\p\chi}{\p
z}\Big)(D_\g\chi)^{-3}\\0 &
(D_\g\chi)^{-1}\end{array}\right)\left(\begin{array}{c}\frac{\p}{\p
z}\\D_\g\end{array}\right)
\ee
Plugging in the parameterization (\ref{scfparam}), and looking at the
pull-pack of the vector field $l(X)$, one finds the induced transformation law
for $X$
\be\label{xxmfn}
X(w,\chi) \mapsto (D_\g\chi)^{-2}(X\circ\Phi)(z,\g )=: \hat{X}(z,\g )
\ee
showing that $X$ can in fact be identified with components of sections of the locally rank one sheaf
$\omega^{-1}$\cite{nagi}\cite{nelson2}, where $\omega = dz + \g d\g$. Thus, in local
co-ordinates, it reads as the `tensor' $\mathbf{X}=X\omega^{-1}$, yielding the
transformation law
\be
\Phi^*(X(w,\chi )\omega^{-1}) =
(X\circ\Phi)(z,\g)\omega^{-1}(D_\g\chi)^{-2} = \hat{X}(z,\g)\omega^{-1}
\ee
A bracket can be
introduced on $\Gamma(\omega^{-1})$ \cite{kac3}as
\be
\co{X_{(1)}}{X_{(2)}}=-2X_{(1)}X_{(2)}' + 2X_{(1)}'X_{(2)} - (-1)^{X_{(1)}}(DX_{(1)})(DX_{(2)})
\ee
The bracket is graded antisymmetric and obeys the graded
Jacobi identity, and hence defines a lie graded algebra structure. The map $l:\Gamma(\omega^{-1})\rightarrow
\Gamma(\mathcal{D}^1\mathcal{A}_1)$ is then a lie graded algebra
homomorphism, with the bracket on vector fields given by the usual lie
bracket. (\ref{n1centr}) can be rewritten as
\be
c(X_{(1)}, X_{(2)}) = \frac{1}{12\pi i}\oint_0 dzd\g \Big(
(DX_{(1)}'')X_{(2)} - (-1)^{X_{(1)}X_{(2)}}(DX_{(2)}'')X_{(1)}\Big)
\ee
The integration `measure', transforms, according to the Berezinian as
\be
\Phi^*\Big[ dw \frac{\p}{\p\chi}\Big] = \Big[ dz \frac{\p}{\p\g}\Big](D_\g\chi) 
\ee
Knowing how $X$ transforms from (\ref{xxmfn}), one can compute
\be\label{n1centrxmfn}
&&\Phi^*\Big((D_\chi X_{(1)}'')X_{(2)} -
(-1)^{X_{(1)}X_{(2)}}(D_\chi X_{(2)}'')X_{(1)}\Big) =\nonumber\\&&\qquad\qquad
(D_\g \chi)^{-1}\Big( (D_\g \hat{X}_{(1)}'')\hat{X}_{(2)} -
(-1)^{X_{(1)}X_{(2)}}(D_\g \hat{X}_{(2)}'')\hat{X}_{(1)} +
2\{\chi,\g\}\co{\hat{X}_{(1)}}{\hat{X}_{(2)}} \nonumber\\&&
\qquad\qquad + D_\g \big(\{\chi,\g\}((D_\g \hat{X}_{(1)}'')\hat{X}_{(2)} -
(-1)^{X_{(1)}X_{(2)}}(D_\g \hat{X}_{(2)}'')\hat{X}_{(1)}\big)\Big)
\ee
where the primes on the left hand side are derivatives with respect to
$w$, and on the right hand side with respect to $z$. Also, $(-1)^{X}\hat{X} = (-1)^{\hat{X}}\hat{X}$ on virtue of
$D_\g\chi$ being even and
\be\label{n1schwarz}
\{\chi,\g \} = \frac{\chi ''}{D_\g\chi} - 2\frac{\chi 'D_\g\chi
'}{(D_\g\chi )^2}
\ee
is the $N=1$ Schwarzian\cite{ks}. Notice that the last term in
(\ref{n1centrxmfn}) is a total derivative in $D_\g$, and hence will
vanish under the integral. Hence, it is most useful to look at
sections of $dz\otimes\frac{\p}{\p\g}$ modulo exact derivatives. More
explicitly,
\be
\oint_{0}dzd\g (f_0 + f_1\g) = (-1)^{f_1}\oint_{0} dz f_1
\ee
and modulo exact derivatives means that that if $f_1$ has an
antiderivative, then $(f_0 + f_1\g)\sim 0$. In particular, this means
\be
f_0 + f_1\g = f_0 + F_1'\g = D_\g ( (-1)^{F_1}F_1 + (-1)^{f_0}f_0\g )\sim 0
\ee
as required. 

Given
$\Phi :(z,\g ) \mapsto (w,\chi )$, a contravariant map can be defined
\be
&&U_{\chi, \g} : \left(\begin{array}{c} \frac{c}{6}( (D_\chi X_{(1)}'')X_{(2)} -
(-1)^{X_{(1)}X_{(2)}}(D_\chi X_{(2)}'')X_{(1)})\\ 
\co{X_{(1)}}{X_{(2)}}\end{array}\right) \mapsto\nonumber\\&&\qquad\qquad\qquad
\left(\begin{array}{c} \frac{c}{6}( (D_\g \hat{X}_{(1)}'')\hat{X}_{(2)} -
(-1)^{X_{(1)}X_{(2)}}(D_\g \hat{X}_{(2)}'')\hat{X}_{(1)}) \\ 
\co{\hat{X}_{(1)}}{\hat{X}_{(2)}}\end{array}\right)
\ee
by
\small
\be
&& U_{\chi, \g}\left(\begin{array}{c} \frac{c}{6}((D_\chi X_{(1)}'')X_{(2)} -
(-1)^{X_{(1)}X_{(2)}}(D_\chi X_{(2)}'')X_{(1)})\\ 
\co{X_{(1)}}{X_{(2)}}\end{array}\right) = \\ &&\left(\begin{array}{cc}(D_\g\chi) & -\frac{c}{3}\{\chi, \g\}
(D_\g\chi)^{-2}\\ 0 & (D_\g\chi)^{-2} \end{array}\right)\Phi^*\left(\begin{array}{c}\frac{c}{6}((D_\chi X_{(1)}'')X_{(2)} -
(-1)^{X_{(1)}X_{(2)}}(D_\chi X_{(2)}'')X_{(1)})\\ \co{X_{(1)}}{X_{(2)}} \end{array}\right)\nonumber
\ee
\normalsize
recalling that the first term is only defined up to exact
derivatives. Given, in addition, a map $\Psi : (w,\chi )\mapsto (u,\rho )$, and
using the property of the Schwarzian $\{\rho, \g\} =
\{\rho,\chi\}(D_\g\chi)^3 + \{\chi,\g\}$ which can be deduced from (\ref{n1schwarz}), one can show
\be\label{restr}
U_{\chi, \g}\circ U_{\rho, \chi} = U_{\rho, \g}
\ee
Associating the necessary open sets to the maps $\Phi$ and
$\Psi$ then realises (\ref{restr}) as the requirement on restriction
maps. This then, locally, represents a nontrivial extension of
\be
\left[ dz\frac{\p}{\p\g}\right]\oplus\omega^{-1}
\ee
modulo exact derivatives in the first slot and for $c\neq 0$. If
$c=0$, the extension becomes trivial. A graded lie bracket can then by
defined on sections of this extension as
\be
\left[ \left(\begin{array}{c}\lambda\\X_{(1)}\end{array}\right),
\left(\begin{array}{c}\mu\\X_{(2)}\end{array}\right)\right] =
\left(\begin{array}{c} \frac{c}{6}((D_\chi X_{(1)}'')X_{(2)} -
(-1)^{X_{(1)}}X_{(1)}(D_\chi X_{(2)}''))\\ 
\co{X_{(1)}}{X_{(2)}}\end{array}\right)
\ee
where the grade is given by the lower component. Note that, although
it looks like the top component has a different parity to the bottom
component, after performing the $\int dz d\g$ integral to get the
central charge, it will have the same parity. The graded Jacobi identity is
already satisfied by the lower component, it remains to verify the
upper component.
\be
(-1)^{X_{(1)}X_{(3)}}\left[\left[\left(\begin{array}{c}\lambda\\X_{(1)}\end{array}\right),
\left(\begin{array}{c}\mu\\X_{(2)}\end{array}\right)\right],
\left(\begin{array}{c}\sigma\\X_{(3)}\end{array}\right) \right] +
\textrm{cyclic} = \left(\begin{array}{c}U\\ 0\end{array}\right)
\ee
has top component
\small
\be
&&U = (-1)^{X_{(1)}X_{(3)}}\Big(\Big( -(DX_{(1)})X_{(2)}''' +
(-1)^{X_{(1)}}X_{(1)}'''(DX_{(2)}) -
3(-1)^{X_{(1)}}X_{(1)}'(DX_{(2)}'') \nonumber\\&&+ 3(DX_1'')X_2' -
2(-1)^{X_{(1)}}X_1(DX_{2}''') +2(DX_{(1)}''')X_{(2)}\Big) X_3 +
(-1)^{X_{(1)}+X_{(2)}}\Big( 2X_{(1)}X_{(2)}' \nonumber\\&&-
2X_{(1)}'X_{(2)} + (-1)^{X_{(1)}}(DX_{(1)}DX_{(2)})\Big)\Big) (DX_3'')
+ \textrm{cyclic}\nonumber\\&&\nonumber\\&&
= (-1)^{X_{(1)}X_{(3)}}\p\Big( 2(DX_{(1)}')X_{(2)}'X_{(3)} -
2(DX_{(1)}')X_{(2)}X_{(3)}' + 2(-1)^{(X_1)}X_{(1)}(DX_{(2)}')X_{(3)}'\nonumber\\&&
- 2(-1)X_{(1)}(DX_{(2)}')X_{(3)} +
2(-1)^{X_{(1)}+X_{(2)}}X_{(1)}'X_{(2)}(DX_{(3)}') -\nonumber\\&&
2(-1)^{X_{(1)}+X_{(2)}}X_{(1)}X_{(2)}'(DX_{(3)}') -
(-1)^{X_{(2)}}\p((DX_{(1)})(DX_{(2)})(DX_{(3)}))\Big) +\nonumber\\&&
(-1)^{X_{(1)}X_{(3)}}D\Big( 2(-1)^{X_{(1)}}(DX_{(1)}') -
2(-1)^{X_{(1)}+X_{(2)}}(DX_{(1)}')X_{(2)}(DX_{(3)}') +\nonumber\\&&
2(-1)^{X_{(2)}}X_{(1)}(DX_{(2)}')(DX_{(3)}') +
(-1)^{X_{(1)}}(DX_{(1)}'')(DX_{(2)})X_{(3)} -\nonumber\\&&
(-1)^{X_{(1)}+X_{(2)}}(DX_{(1)}'')X_{(2)}(DX_{(3)})
+(-1)^{X_{(1)}}(DX_{(1)})(DX_{(2)}'')X_{(3)} +\nonumber\\&&
(-1)^{X_{(2)}}X_{(1)}(DX_{(2)}'')(DX_{(3)}) +
(-1)^{X_{(2)}}X_{(1)}(DX_{(2)})(DX_{(3)}'') -\nonumber\\&&
(-1)^{X_{(1)}+X_{(2)}}(DX_{(1)})X_{(2)}(DX_{(3)}'')\Big)\sim 0
\ee
\normalsize
where the equivalence follows since the term is a total derivative
(recall $\p = D^2$),
and the bracket obeys the graded Jacobi identity. Hence, the
algebra of conformal vector fields admits an extension on the graded Riemann
sphere to accommodate the central charge, which realises a map of the
form (\ref{conj}) using (\ref{n1centr}). The construction precisely
mirrors that done for the bosonic case in \cite{witten}. Given the
list of Schwarzians in \cite{ks}, it should be possible to extend the
construction to all $N\leq 4$.

\section{Conclusions}

If one wishes to study all of the conformal symmetries on an $N=4$
graded Riemann sphere at the quantum level, and not neglect any
symmetries, then one is forced to look at adding non-central
extensions to the algebra. If the $u_0$ symmetry is neglected, then
all of the extensions are central. To the author's knowledge, this has
not been observed before, and the author is unaware of the $N=4$
algebra given by (\ref{commrelns}) having appeared previously in the
literature. It was found that this extension was completely consistent
with the usual CFT treatments of the quantum theory, i.e. with the
Operator Product Expansions of the $N=4$ theory, and in fact explained
why the logarithms appeared in the OPE. The effect on the
representation theory was discussed, as was the description of the
Gelfand-Fuchs extensions for $N=1\ldots 4$, given by the superfield formalism. There,
it was found that the extra extension explained the appearance of the
indefinite integral inside the contour integral in (\ref{cpgf}). It
was then described for the $N=1$ case how the central extension
arises from the geometry of an $N=1$ graded Riemann sphere, and how it
might be expected that the central extensions for the higher $N$ might
be obtained.

As well as exploiting graded Riemann sphere geometry to describe
extensions of Superconformal algebras, these calculations also shed
some light on how one might try to treat non-central extensions in a
CFT. In this case, the Field Theory exhibited an, albeit mild,
logarithmic behaviour, not unrelated to the manner in which zero modes
are modified in \cite{flo2}. For higher $N$, the requirement of having
a Virasoro subalgebra with non-zero central extension could force non-central extensions
to appear that enrich the zero mode structure. $N=5$ would have a
graded odd weight $-\frac{1}{2}$ field whose zero mode structure might give unavoidable
Jordan cells in the representation theory. In the $N=4$ case, the modified zero mode
structure was reflected by the appearance of indefinite integrals in
the Gelfand-Fuchs extensions.

\section{Acknowledgements}

The author would like to thank Dr A. Taormina, Dr J. Evans and Prof
P. Sorba for helpful discussions.


\newpage

\begin{thebibliography}{99}
\bibitem{flo} Flohr, M., ``Bits and Pieces in Logarithmic Conformal
Field Theory'' Int. J. Mod. Phys. A \textbf{18}, 4497-4592 (2003)
\bibitem{flo2} Flohr, M., ``Ghost system revisited: Modified Virasoro
Generators and Logarithmic Conformal Field Theories'' JHEP \textbf{1}
20-38 (2003)
\bibitem{kent} Goddard, P., Kent, A., Olive, D., ``Virasoro Algebras
and Coset Space Models'' Phys. Lett \textbf{B152} 88-102 (1985)
\bibitem{groz} Grozman, P., Leites, D., Shchepochkina, I., ``Lie
Superalgebras of String Theories'' hep-th/9702120
\bibitem{ivan} Ivanov, E., Krivonos, S., Leviant, V. ``$N=3$ and $N=4$
Superconformal WZNW sigma models in Superspace II - the $N=4$ case''
Int. J. Mod. Phys. A \textbf{7} 287-316 (1992)
\bibitem{kac3} Kac, V., van de Leur, J., ``On Classification of
Superconformal Algebras'' Proceedings, Strings (1988)
\bibitem{nagi} Nagi, J., ``Superconformal Primary Fields on a Graded
Riemann Sphere'' J. Math. Phys. \textbf{45}, 2492-2514 (2004)
\bibitem{ks} Schoutens, K., ``$O(N)$-extended Superconformal Field
Theory in Superspace'' Nuc. Phys. \textbf{B295}, 634-652 (1988)
\bibitem{nelson2} Nelson, P., ``Lectures on Supermanifolds and
Strings'',  Published in Providence: TASI 88:0997-1074
(QCD161:T45:1988:V.2)
\bibitem{tao1} Petersen J., Taormina A. ``Characters of the $N=4$
Superconformal Algebra with two Central Extensions''
Nuc. Phys. \textbf{B331}, 556-572 (1990)
\bibitem{tao2} Petersen J., Taormina A. ``Characters of the $N=4$
Superconformal Algebra with two Central Extensions 2: Massless Representations''
Nuc. Phys. \textbf{B333}, 833-854 (1990)
\bibitem{vanp} Sevrin A., Troost W., Van Proeyen A., ``Superconformal
Algebras in Two Dimensions with $N=4$'' Phys. Lett. \textbf{B208}, 447-450 (1988)
\bibitem{witten} Witten, E., ``Quantum Field Theory, Grassmannians, and
Algebraic Curves'' Commun. Math. Phys. \textbf{113}, 529 (1988)
\bibitem{ft1} if $ U(z)\ket{0}$ is to be regular at $z=0$,
  then $V_0$ must annihilate $\ket{0}$
\bibitem{ft2} e.g. in the $N=0$
case $\frac{1}{\pi i}\oint_0 dz X_{(i)}\mathbb{T} = \frac{1}{2\pi
  i}\oint_0 dz l_{(i)}L(z)$ . The vector $l_n$ is parameterized by
$l_{(i)}=z^{n+1}$. Then $\frac{1}{2\pi i}\oint_0 dz l_{(i)}L(z) =
\frac{1}{2\pi i}\oint_0 z^{n+1}\sum_m L_mz^{-m-2} = L_n$
\bibitem{ft3}more generally, $(-1)^{\Pi_iX_{(i)}}\Pi_iX_{(i)} =
\sum_{n_i\in\{ 0,1 \} }(-1)^{\Pi_iX_{(i)n_i}}\Pi_iX_{(i)n_i}$
\end{thebibliography}
\end{document}